\documentclass[a4paper,12pt]{article}
\usepackage{graphicx,amssymb}

\newcommand{\bea}{\begin{eqnarray}}
\newcommand{\ena}{\end{eqnarray}}
\newcommand{\vs}[1]{\vspace{#1 mm}}
\newcommand{\hs}[1]{\hspace{#1 mm}}
\renewcommand{\a}{\alpha}
\renewcommand{\b}{\beta}
\renewcommand{\c}{\gamma}
\renewcommand{\d}{\delta}
\newcommand{\e}{\epsilon}
\newcommand{\s}{\sigma}

\def\bbox{{\,\lower0.9pt\vbox{\hrule \hbox{\vrule height 0.2 cm
\hskip 0.2 cm \vrule height 0.2 cm}\hrule}\,}}
\newcommand{\dsl}{\pa \kern-0.5em /}

\newcommand{\shalf}{\frac{1}{2}}
\newcommand{\pa}{\partial}
\newcommand{\nn}{\nonumber\\}
\newcommand{\p}[1]{(\ref{#1})}

\begin{document}

\renewcommand{\thefootnote}{\fnsymbol{footnote}}
\begin{titlepage}

\setcounter{page}{0}
\begin{flushright}
OU-HET 428 \\
hep-th/0301095
\end{flushright}

\vs{10}
\begin{center}
{\Large\bf Intersection Rules for S-Branes}
\vs{30}

{\large
Nobuyoshi Ohta\footnote{e-mail address: ohta@phys.sci.osaka-u.ac.jp}}\\
\vs{10}
{\em Department of Physics, Osaka University,
Toyonaka, Osaka 560-0043, Japan}
\end{center}
\vs{15}
\centerline{{\bf{Abstract}}}
\vs{5}

We give a model-independent derivation of general intersecting rules for
spacelike branes (S-branes) in arbitrary dimensions $d$. This is achieved by
directly solving bosonic field equations for supergravity coupled to
a dilaton and antisymmetric tensor fields with minimal ans\"{a}tze.
We compare the results with those in eleven-dimensional supergravity
and other solutions.

\end{titlepage}
\newpage
\renewcommand{\thefootnote}{\arabic{footnote}}
\setcounter{footnote}{0}
\setcounter{page}{2}

There has been much interest in time-dependent and spacelike brane solutions
(S-branes) of supergravities in eleven and ten dimensions because of
its possible connection with tachyon condensations and dS/CFT
correspondence~\cite{GS,S}. These theories are the low-energy
limits of the string theories and supposedly unifying M-theory of strings.
Time-dependent solutions are investigated rather recently and not much is
known on these solutions. It is thus important to better understand these
classical $p$-brane solutions.

The single $p$-brane solutions (S$p$-branes) have been discussed in
refs.~\cite{GS,CGG,KMP,R} for low-energy effective supergravities (see also
refs.~\cite{LMP,H} for related solutions). Following the usual convention,
S$p$-branes are used for those with $(p+1)$-dimensional Euclidean world-volume.
It has then been noted that the more general solutions can be understood
as intersecting ones of these fundamental $p$-branes~\cite{DK,V} and the
rules how the branes intersect with each other are given in analogy to
the usual branes~\cite{PT}--\cite{NO}. Although the ``rules'' are consistent
with most of the known solutions, it is not clear if there are any other
solutions than those given by these rules. The questions we would like to
ask here are how general these rules are and how severely they restrict
the solutions for supergravities in $d=11$ and lower dimensions.

A systematic approach to formulating the rules for the way how they can
intersect has been derived for the usual branes in \cite{AEH,NO}.
The purpose of this note is to extend this work to the S-branes and
clarify what ans\"{a}tze are really necessary. In particular,
we derive the intersection rules from the general approach.
We show that the rules are simple consequences of the field equations,
which can be easily integrated and the consistency of the solutions
reduces the problem of solving the field equations to an algebraic one.

The results of our analysis turn out to be consistent with the
superposition rules in ref.~\cite{DK} for $d=11$ supergravity, but our results
apply to more general supergravity coupled to a dilaton and antisymmetric
tensors. We show that the requirement that the fields for the each brane be
independent is sufficient to give the solutions and intersection rules.

Let us start with the general action for gravity coupled to a dilaton
$\phi$ and $m$ different $n_A$-form field strengths:
\bea
I = \frac{1}{16 \pi G_d} \int d^d x \sqrt{-g} \left[
R - \shalf (\pa \phi)^2 - \sum_{A=1}^m \frac{1}{2 n_A!} e^{a_A \phi}
F_{n_A}^2 \right].
\label{act}
\ena
This action describes the bosonic part of $d=11$ or $d=10$ supergravities;
we simply drop $\phi$ and put $a_A=0$ and $n_A=4$ for $d=11$, whereas we
set $a_A=-1$ for the NS-NS 3-form and $a_A=\shalf(5-n_A)$ for forms coming
from the R-R sector.\footnote{There may be Chern-Simons terms in the action,
but they are irrelevant in our following solutions.} To describe more
general supergravities in lower dimensions, we should include several scalars,
but for simplicity we disregard this complication in this paper.

{}From the action (\ref{act}), one derives the field equations
$$
R_{\mu\nu} = \shalf \pa_\mu \phi \pa_\nu \phi + \sum_{A} \frac{1}{2 n_A!}
e^{a_A \phi} \left[ n_A \left( F_{n_A}^2 \right)_{\mu\nu}
- \frac{n_A -1}{d-2} F_{n_A}^2 g_{\mu\nu} \right],
$$
$$
\bbox \phi = \sum_{A} \frac{a_A}{2 n_A!} e^{a_A \phi} F_{n_A}^2,
$$
$$
\pa_{\mu_1} \left( \sqrt{- g} e^{a_A \phi} F^{\mu_1 \cdots \mu_{n_A}} \right)
= 0,
$$ \vs{-10}
\bea
\pa _{[\mu} F_{\mu_1 \cdots \mu_{n_A}]} = 0.
\label{fe}
\ena
The last equations are the Bianchi identities.

We take the following metric for our system:
\bea
ds_d^2 = -e^{2u_0} dt^2 + \sum_{\a=1}^{p} e^{2 u_\a} dy_\a^2
+ e^{2B} d\Sigma_{k,\s}^2,
\label{met}
\ena
where $d=p+k+1$, the coordinates $y_\a, (\a=1,\ldots, p)$ parametrize the
$p$-dimensional world-volume directions and the remaining coordinates of
the $d$-dimensional spacetime are the time $t$ and coordinates on
$k$-dimensional spherical ($\s=+1$), flat ($\s=0$) or hyperbolic
($\s=-1$) spaces, whose line elements are $d\Sigma_{k,\s}^2$.
Since we are interested in time-dependent solutions, all the functions
appearing in the metrics as well as dilaton $\phi$ are assumed to depend
only on the time $t$. The Ricci tensors for the metric~\p{met} are
\bea
R_{00} &=& - \sum_{\a=1}^p [ \ddot u_\a + (\dot u_\a)^2 -\dot u_\a \dot u_0]
- k(\ddot B + \dot B^2 - \dot B \dot u_0 ), \nn
R_{\a\b} &=& e^{2(u_\a-u_0)} [\ddot u_\a -\dot u_\a \dot u_0
+ \sum_{\c=1}^p \dot u_\c \dot u_\a + k \dot B \dot u_\a]\d_{\a\b}, \nn
R_{ab} &=& \Big\{ e^{2(B-u_0)}[\ddot B + k \dot B^2 -\dot B \dot u_0
+ \sum_{\a=1}^p \dot u_\a \dot B] + \s (k-1)\Big\}\bar g_{ab},
\label{ricci}
\ena
where $\bar g_{ab}$ is the metric for the hypersurface $\Sigma_{k,\s}$.
Here and in what follows, a dot denotes a derivative with
respect to $t$.

For the field strengths, we take the most general ones consistent
with the field equations and Bianchi identities.
The value for an electrically charged S$q$-brane (whose world-volume
is $(q+1)$-dimensional) is given by
\bea
F_{t \a_1 \cdots \a_{q+1}} = \e_{\a_1 \cdots \a_{q+1}} \dot E, \hs{3}
(n_A = q+2),
\label{ele}
\ena
where $\a_1, \cdots ,\a_{q+1}$ stand for the tangential direction to
the S$q$-brane. The magnetic case is given by
\bea
F^{\a_{q+2} \cdots \a_p a_1 \cdots a_{k}} = \frac{1}{\sqrt{-g}}
e^{-a\phi} \e^{t \a_{q+2} \cdots \a_p a_1 \cdots a_k} \dot{\tilde E},
\hs{3} (n_A = d-q-2)
\label{mag}
\ena
where $a_1, \cdots, a_{k}$ denote the coordinates of the $k$-dimensional
hypersurface $\Sigma_{k,\s}$.
The functions $E$ and $\tilde E$ are again assumed to depend only on $t$.

The electric field (\ref{ele}) trivially satisfies the Bianchi
identities but the field equations are nontrivial. On the other hand, the
field equations are trivial but the Bianchi identities are nontrivial
for the magnetic field (\ref{mag}).

We will solve the field eqs.~(\ref{fe}) with the simplifying ansatz
\bea
- u_0 + \sum_{\a=1}^p u_\a + k B = 0,
\label{ans}
\ena
which simplifies the field equations~(\ref{fe}) considerably.
For both cases of electric~(\ref{ele}) and magnetic~(\ref{mag})
fields, we find that the field eqs.~(\ref{fe}) are cast into
\bea
&& - \ddot u_0 + (\dot u_0)^2 - \sum_{\a=1}^p (\dot u_\a)^2 - k \dot B^2
= \frac{1}{2} \dot\phi^2 + \sum_{A} \frac{d-q_A-3}{2(d-2)} S_A ({\dot E_A})^2,
\label{1}
\\
&& \ddot u_\a = - \sum_{A} \frac{\d_A^{(\a)}}{2(d-2)} S_A ({\dot E_A})^2,
\hs{3} (\a=1,\cdots,p),
\label{2}
\\
&& \ddot B + \s (k-1) e^{2u_0-2B} = \sum_{A} \frac{q_A+1}{2(d-2)}
S_A ({\dot E_A})^2,
\label{3}
\\
&& \ddot \phi = \sum_{A} \frac{\e_A a_A}{2} S_A ({\dot E_A})^2,
\label{4}
\\
&& \left( S_A {\dot E_A} \right)^. = 0,
\label{5}
\ena
where $A$ denotes the kinds of $q_A$-branes and we have defined
\bea
S_A \equiv \exp \left( \e_A a_A \phi - 2 \sum_{\a \in q_A} u_\a \right),
\label{6}
\ena
and
\bea
\d_A^{(\a)} = \left\{ \begin{array}{l}
d - q_A - 3 \\
- (q_A+1)
\end{array}
\right.
\hs{5}
{\rm for} \hs{3}
\left\{
\begin{array}{l}
y_\a \hs{3} {\rm belonging \hs{2} to} \hs{2} q_A{\rm -brane} \\
{\rm otherwise}
\end{array},
\right.
\ena
and $\e_A= +1 (-1)$ corresponds to electric (magnetic) fields.
For magnetic case we have dropped the tilde from $E_A$. Equations~\p{1},
\p{2} and \p{3} are the $00, \a\a$ and $ab$ components of the Einstein
equation in \p{fe}, respectively. The last one is the field equation for
the field strengths of the electric fields and/or Bianchi identity
for the magnetic ones. It is remarkable that both the electric and magnetic
cases can be treated simultaneously just by using the sign $\e_A$. This is
because the original system~\p{act} has the S-duality symmetry under
\bea
g_{\mu\nu} \to g_{\mu\nu}, \quad
F_{n_A} \to e^{-a_A \phi}*\! F_{n_A}, \quad
\phi \to - \phi.
\label{sdual}
\ena

{}From eq.~(\ref{5}) one finds
\bea
S_A \dot E_A = c_A,
\label{const}
\ena
where $c_A$ is a constant.
With the help of eq.~\p{const}, we find that eqs.~\p{2} and \p{4} give
\bea
\dot u_\a &=& - \sum_{A} \frac{\d_{A}^{(\a)}}{2(d-2)} c_A E_A + c_\a, \nn
\dot \phi &=& \sum_{A} \frac{\e_A a_A}{2} c_A E_A + c_\phi,
\label{fint}
\ena
where $c_\a$ and $c_\phi$ are integration constants.
Let us next define
\bea
g(t) = (u_0-B)/(k-1).
\ena
We find from \p{ans}
\bea
B = g - \frac{1}{k-1}\sum_{\a=1}^p u_\a, \qquad
u_0 = kg - \frac{1}{k-1}\sum_{\a=1}^p u_\a,
\ena
Using \p{fint}, we get
\bea
\label{bdot}
\dot B = \dot g + \sum_{A} \frac{q_A+1}{2(d-2)} c_A E_A
- \frac{1}{k-1}\sum_{\a=1}^p c_\a, \\
\dot u_0= k \dot g + \sum_{A} \frac{q_A+1}{2(d-2)} c_A E_A
- \frac{1}{k-1}\sum_{\a=1}^p c_\a,
\label{u0dot}
\ena
Substituting \p{const} and \p{bdot} into \p{3}, we obtain
\bea
\ddot g + \s (k-1) e^{2(k-1)g} = 0,
\label{gddot}
\ena
which yields
\bea
\dot g^2 + \s e^{2(k-1)g} = \b^2,
\label{gdot}
\ena
where $\b$ is an integration constant. The solution to eq.~\p{gdot} is
given by
\bea
g(t) = \left\{\begin{array}{ll}
\frac{1}{k-1} \ln \frac{\b}{\cosh[(k-1)\b(t-t_1)]} & :\s=+1, \\
\pm \b(t-t_1) & :\s=0, \\
\frac{1}{k-1} \ln \frac{\b}{\sinh[(k-1)\b(t-t_1)]} & :\s=-1,
\end{array}
\right.
\ena
where $t_1$ is another integration constant.

Substituting eqs.~(\ref{fint}), \p{gddot} and \p{gdot} into
(\ref{1}) yields
\bea
&& (k-1) \left( \sum_{A} \frac{q_A+1}{2(d-2)} c_A E_A
- \frac{1}{k-1}\sum_{\a=1}^p c_\a\right)^2
+ \sum_{\a=1}^p \left( \sum_{A} \frac{\d_{A}^{(\a)}}{2(d-2)} c_A E_A
- c_\a \right)^2 \nn
&& + \; \shalf \left( \sum_{A} \frac{\e_A a_A}{2} c_A E_A + c_\phi \right)^2
+ \sum_{A} \frac{c_A}{2} \dot E_A - k(k-1) \b^2 =0.
\label{sint}
\ena
This equation must be valid for arbitrary functions $E_A$ of $t$.
{}From the $E_A$-independent part of eq.~(\ref{sint}), one finds
\bea
\frac{1}{k-1}\left(\sum_{\a=1}^p c_\a\right)^2
+ \sum_{\a=1}^p c_\a^2 + \shalf c_\phi^2= k(k-1) \b^2.
\label{condconst}
\ena
We can then rewrite eq.~(\ref{sint}) as
\bea
\sum_{A,B} \left[ M_{AB} \frac{c_A}{2} - \d_{AB} \left\{
\left( \frac{1}{E_A}\right)^. + \frac{2\tilde c_A}{E_A}\right\} \right]
\frac{c_B}{2} E_A E_B =0,
\label{tint}
\ena
where
\bea
\label{cond1}
M_{AB} &=& \sum_{\a=1}^p \frac{\d_{A}^{(\a)}\d_{B}^{(\a)}}{(d-2)^2}
+ (k-1) \frac{(q_A+1)(q_B+1)}{(d-2)^2} + \shalf \e_A a_A \e_B a_B, \\
\tilde c_A &=& \sum_{\a\in q_A} c_\a-\frac{1}{2} c_\phi \e_A a_A.
\label{cond11}
\ena
Since $M_{AB}$ is constant, eq.~(\ref{tint}) cannot be satisfied for
arbitrary functions $E_A$ of $t$ unless the second term inside
the square bracket is a constant. Requiring this to be a constant tells
us that the function $E_A$ must satisfy
\bea
\left(\frac{1}{E_A}\right)^. + \frac{2\tilde c_A}{E_A} = \tilde c_A N_A,
\label{har}
\ena
or
\bea
E_A = \frac{e^{\tilde c_A(t-t_A)}}{N_A \cosh \tilde c_A(t-t_A)},
\label{har1}
\ena
where $N_A$ is a normalization factor and $t_A$ is an integration constant.
In this way, the problem reduces to the algebraic equation (\ref{tint})
supplemented by (\ref{har}) without making any assumption other than
(\ref{ans}).

Equation~(\ref{tint}) has two implications if we take
independent functions for the fields $E_A$. In this case,
first putting $A=B$ in eq.~(\ref{tint}), we learn that
\bea
c_A = \frac{2(d-2)\tilde c_A N_A}{\Delta_A},
\label{cond2}
\ena
where
\bea
\Delta_A = (q_A + 1) (d-q_A-3) + \shalf a_A^2 (d-2).
\label{res2}
\ena
By use of eqs.~\p{har1} and \p{cond2}, eqs.~\p{fint}, \p{bdot} and \p{u0dot}
can be integrated with the results
\bea
u_0 &=& kg(t) + \sum_{A} \frac{q_A+1}{\Delta_A} \ln \cosh\tilde c_A(t-t_A)
+c_0 t + c_0', \nn
u_\a &=& - \sum_{A} \frac{\d_{A}^{(\a)}}{\Delta_A} \ln \cosh\tilde c_A(t-t_A)
+ \tilde c_\a t + c_\a', \nn
B &=& g(t) + \sum_{A} \frac{q_A+1}{\Delta_A} \ln \cosh\tilde c_A(t-t_A)
+ c_0 t +c_0', \nn
\phi &=& \sum_{A} \frac{(d-2)\e_A a_A}{\Delta_A} \ln \cosh\tilde c_A(t-t_A)
+\tilde c_\phi t + c_\phi',
\label{res3}
\ena
where $c_\a'$'s are new integration constants and
\bea
&& c_0 = \sum_A \frac{q_A+1}{\Delta_A}\tilde c_A
-\frac{\sum_{\a=1}^p c_\a}{k-1}, \;\;
c_0' = -\frac{\sum_{\a=1}^p c_\a'}{k-1}, \;\;
\tilde c_\a = c_\a - \sum_A \frac{\d_{A}^{(\a)}}{\Delta_A}\tilde c_A, \nn
&& \tilde c_\phi = c_\phi + \sum_A \frac{(d-2)\e_A a_A}{\Delta_A}\tilde c_A.
\ena

To fix the normalization $N_A$, we go back to eq.~(\ref{6}).
Using (\ref{res3}), we find
\bea
S_A = [\cosh\tilde c_A(t-t_A)]^2 e^{\e_Aa_A c_\phi'-2\sum_{\a\in q_A} c_\a'},
\label{ress}
\ena
which, together with (\ref{const}) and (\ref{cond2}), leads to
\bea
N_A = \sqrt{\frac{\Delta_A}{2(d-2)}}
e^{\e_Aa_A c_\phi'/2-\sum_{\a\in q_A} c_\a'}.
\label{norm}
\ena

Our metric and other fields are thus finally given by
\bea
ds_d^2 &=& \prod_A [\cosh\tilde c_A (t-t_A)]^{2 \frac{q_A+1}{\Delta_A}}
\Bigg[ e^{2kg(t)+2c_0 t+2c_0'} \left\{ - dt^2
+ e^{-2(k-1)g(t)}d\Sigma_{k,\s}^2\right\} \nn
&& \hs{20} +\; \sum_{\a=1}^{p} \prod_A [\cosh\tilde c_A(t-t_A)]^{- 2
\frac{\c_A^{(\a)}}{\Delta_A}} e^{2 \tilde c_\a t+2c_\a'} dy_\a^2\Bigg], \nn
E_A &=& \frac{e^{\tilde c_A(t-t_A)}}{N_A \cosh \tilde c_A(t-t_A)},\quad
\tilde c_A = \sum_{\a\in q_A} c_\a-\frac{1}{2} c_\phi \e_A a_A.
\ena
where we have defined
\bea
\c_A^{(\a)} = \left\{ \begin{array}{l}
d-2 \\
0
\end{array}
\right.
\hs{5}
{\rm for} \hs{3}
\left\{
\begin{array}{l}
y_\a \hs{3} {\rm belonging \hs{2} to} \hs{2} q_A{\rm -brane} \\
{\rm otherwise}
\end{array}.
\right.
\ena
These solutions contain $2p+2$ integration constants $c_\a, c_\a'
(\a=1, \cdots,p), c_\phi, c_\phi'$, and $t_1$ and $t_A$ with $\b$
determined by eq.~\p{condconst}. Among these, $c_\a'$ can
be removed by rescaling the coordinates, and $t_1$ by a shift of the time.
Without any preference of the choice of other paremeters, we leave
these as free parameters. Thus the general solutions can be constructed by
the following rules:
(1) All the directions are multiplied by $[\cosh\tilde c_A(t-t_A)]^{2
\frac{q_A+1}{\Delta_A}}$, and in addition,
(2) the overall transverse direction (time and $k$-dimensional space) has
the form $e^{2c_0t}[-e^{2kg(t)}dt^2 + e^{2g(t)}d\Sigma_{k,\s}^2]$ up to
the rescaling of the coordinates,
(3) the coordinates belonging to the brane are multiplied by
$[\cosh\tilde c_A(t-t_A)]^{-2\frac{d-2}{\Delta_A}}$.
When these are specified to $d=11$ supergravity and the integration
constants are chosen appropriately, these give the superposition
rules discussed in ref.~\cite{DK}.

The solution given in ref.~\cite{CGG} is reproduced if we restrict these to
a single S-brane and choose the integration constants as
\bea
&& c_\c = \a - \frac{a}{\chi} c_1, (\c=1,\cdots, p); \;\;
c_\c =-\frac{p}{q-1}\Big(\a-\frac{a}{\chi}c_1\Big), (\c=p+1,\cdots, p+q-k);\nn
&& c_\phi = \frac{d-2}{q-1}a \a + \frac{2p}{\chi}c_1; \;\;
c_\c' = \frac{1}{\chi}\Big(\ln\frac{(d-2)\chi\a^2}{(q-1)b^2} - ac_2\Big),
(\c=1,\cdots, p) ; \nn
&& c_0' = c_b' = c_\c' = -\frac{p}{\chi(q-1)}\Big(\ln\frac{(d-2)\chi\a^2}
{(q-1)b^2} - ac_2\Big), (\c=p+1,\cdots, p+q-k),
\ena
with $q \to p-1, p\to p+q-k, \e=-1$ and $\chi=2p+(d-2)a^2/(q-1)$.\footnote{
Here $c_1$ and $c_2$ on the rhs are those used in ref.~\cite{CGG} and
should not be confused with our $c_\c,(\c=1,2)$.}
Equation \p{condconst} reduces to
\bea
\frac{pc_1^2}{\chi} + \frac{(d-2)\chi\a^2}{2(q-1)}-k(k-1)\b^2=0,
\ena
and the normalization is determined to be
\bea
N_A=\frac{(q-1)b}{2(d-2)\a},
\ena
in complete agreement with ref.~\cite{CGG}.

The second condition following from eqs.~(\ref{tint}) is $M_{AB}=0$ for
$A \neq B$. This leads to the intersection rules for two branes:
If $q_A$-brane and $q_B$-brane intersect over ${\bar q} (\leq q_A, q_B)$
dimensions, this gives
\bea
{\bar q} = \frac{(q_A+1)(q_B+1)}{d-2}-1 - \shalf \e_A a_A \e_B a_B.
\label{int}
\ena
Remember that the world-volume of $q$-branes lies in $(q+1)$-dimensional
space and not in time. For eleven-dimensional supergravity, we have electric
S2-branes, magnetic S5-branes and no dilaton $a_A=0$. The rule \p{int}
tells us that S2-brane can intersect with S2-brane over a `0-brane'
$(\bar q=0)$ (which actually lives in 1-dimensional space) and with S5-brane
over a `string' $(\bar q=1)$ (2-dimensional space), and S5-brane can intersect
with S5-brane over `3-brane' $(\bar q=3)$ (4-dimensional space), again
in agreement with refs.~\cite{DK}. In particular, our results show that
there is no other intersecting solution as long as we treat the functions
$E_A$ with different index $A$ as independent. If this condition is
relaxed, there may be other solutions.

For all the spacelike D$q$-brane solutions in type II superstrings, we find
\bea
\e a =\frac{3-q}{2},
\ena
which tells us that the intersection rule is
\bea
\bar q = \frac{q_A+q_B}{2}-2.
\ena

These solutions do not preserve any supersymmetry. In fact, they are
supposed to correspond to branes with Dirichlet boundary conditions in
the time direction, and hence describe configurations which exist
only for a fixed time. These also contain singularity somewhere in time.
Closer examination of their properties would be quite interesting.

We note that our derivation is a simple generalization of the general
method developed in ref.~\cite{NO}. It is quite satisfying to see
that this is so useful method. There are also an important class of
time-dependent solutions called null-branes~\cite{KR,FS}, which preserve
supersymmetry. It would be also interesting to apply our method to
these solutions.

To summarize, we have given quite a general model-independent derivation
of the superposition rules in arbitrary dimensions.
The intersection rules simply follow from the field equations if we require
that the functions $E_A$ with different index $A$ be independent.
In all cases, the algebraic eq.~(\ref{tint}) (together with (\ref{har}))
must be satisfied, and this equation should be most useful to examine
possible solutions.
We hope to discuss various properties of these solutions using the hints
from dualities implied by underlying string dynamics elsewhere.

\section*{Acknowledgement}
This work was supported in part by Grants-in-Aid for Scientific Research
Nos. 12640270 and 02041.

\newcommand{\NP}[1]{Nucl.\ Phys.\ {\bf #1}}
\newcommand{\PL}[1]{Phys.\ Lett.\ {\bf #1}}
\newcommand{\CQG}[1]{Class.\ Quant.\ Grav.\ {\bf #1}}
\newcommand{\CMP}[1]{Comm.\ Math.\ Phys.\ {\bf #1}}
\newcommand{\IJMP}[1]{Int.\ Jour.\ Mod.\ Phys.\ {\bf #1}}
\newcommand{\JHEP}[1]{JHEP\ {\bf #1}}
\newcommand{\PR}[1]{Phys.\ Rev.\ {\bf #1}}
\newcommand{\PRL}[1]{Phys.\ Rev.\ Lett.\ {\bf #1}}
\newcommand{\PRE}[1]{Phys.\ Rep.\ {\bf #1}}
\newcommand{\PTP}[1]{Prog.\ Theor.\ Phys.\ {\bf #1}}
\newcommand{\PTPS}[1]{Prog.\ Theor.\ Phys.\ Suppl.\ {\bf #1}}
\newcommand{\MPL}[1]{Mod.\ Phys.\ Lett.\ {\bf #1}}
\newcommand{\JP}[1]{Jour.\ Phys.\ {\bf #1}}

\end{document}